\documentclass[showpacs,showkeys,preprintnumbers,pre,amsmath,amssymb,superscriptaddress]{revtex4}
\pdfoutput=1

\usepackage{graphicx}
\usepackage{hyperref}
\usepackage{graphics}
\usepackage{epsfig}
\usepackage{bm}
\usepackage{epic,eepic}
\usepackage{epsfig}
\usepackage{pifont}

\usepackage{caption}
\usepackage{subfigure}

\usepackage{xspace}
\usepackage{graphicx,color}
\usepackage{amsmath,amssymb,amsfonts,mathrsfs}
\usepackage{url}
\usepackage[normalem]{ulem}
\usepackage{booktabs}
\usepackage{float}

\usepackage[toc,page]{appendix}
\usepackage[capitalize]{cleveref}
\usepackage{mathtools}
\usepackage[normalem]{ulem}

\providecommand{\st}[1]{{}^{_{\text{#1}}}\!}% tiny texted subscript
\newcommand{\be}{\begin{equation}}
\newcommand{\ee}{\end{equation}}
\newcommand{\ben}{\begin{equation*}}
\newcommand{\een}{\end{equation*}}

\newcommand{\bv}[1]{\mathbf{#1}}% the vector #1

\newcommand{\vv}{\bv{v}}% hydrodynamical velocity
\newcommand{\delF}[1]{\textcolor{blue}{}}

\newcommand{\rhop}{\rho_{\mathrm{p}}}

\newcommand{\uconf}{U^{\mathrm{(z)}}_{\mathrm{conf}}}
\newcommand{\uo}{U^{\mathrm{(z)}}_{\mathrm{unconf}}}

\newcommand{\kbt}{{\mathrm{k_B T}}}
\newcommand*{\fg}{\mathrm{g}}
\newcommand*{\fp}{\mathrm{F_p}}

\newcommand*{\gconf}{\gamma_{\mathrm{conf}}}
\newcommand*{\go}{\gamma_{\mathrm{unconf}}}
\newcommand*{\cm}{c_m}
\newcommand*{\sigp}{\sigma_{\mathrm{p}}}
\newcommand*{\massp}{m_{\mathrm{p}}}

\newcommand\R[1]{\textcolor{black}{#1}}

\begin{document}

\title{A lattice Boltzmann study of particle settling in a fluctuating multicomponent fluid under confinement}

\author{Xiao Xue}
\email{xiaox@chalmers.se}
\affiliation{Department of Physics  and J.M. Burgerscentrum, Eindhoven University of Technology, 5600 MB Eindhoven, the Netherlands.}
\affiliation{Department of Physics \& INFN, University of Rome ``Tor Vergata'', Via della Ricerca Scientifica 1, 00133, Rome, Italy.}
\affiliation{Department of Mechanics and Maritime Sciences, Division of Fluid Dynamics, Chalmers University of Technology, 41296, G{\"o}teborg, Sweden}

\author{Luca Biferale}
\email{biferale@roma2.infn.it}
\affiliation{Department of Physics \& INFN, University of Rome ``Tor Vergata'', Via della Ricerca Scientifica 1, 00133, Rome, Italy.}

\author{Mauro Sbragaglia}
\email{sbragaglia@roma2.infn.it}
\affiliation{Department of Physics \& INFN, University of Rome ``Tor Vergata'', Via della Ricerca Scientifica 1, 00133, Rome, Italy.}

\author{Federico Toschi}
\email{F.Toschi@tue.nl}
\affiliation{Departments of Physics and of Mathematics and Computer Science and J.M. Burgerscentrum, Eindhoven University of Technology, 5600 MB Eindhoven, the Netherlands.}
\affiliation{ Istituto per le Applicazioni del Calcolo CNR, Via dei Taurini 19, 00185 Rome, Italy.}

%%%%%%%%%%%%%%%%%%%%%%%%%%%%%%%%%%%%%%%%%%%%%%%%%%%%%%%%%%%%%%%%%%%%%%%%%%%%%%%%%%%%%%%%%%%%%%%%%%%%%%%%%
% Abstract
%%%%%%%%%%%%%%%%%%%%%%%%%%%%%%%%%%%%%%%%%%%%%%%%%%%%%%%%%%%%%%%%%%%%%%%%%%%%%%%%%%%%%%%%%%%%%%%%%%%%%%%%%
\begin{abstract}

We present mesoscale numerical simulations based on the coupling of the fluctuating lattice Boltzmann method (FLBM) for multicomponent systems with a wetted finite-size particle model. This newly coupled methodologies are used to study the motion of a spherical particle driven by a constant body force in a confined channel with a fixed square cross-section. The channel is filled with a mixture of two liquids under the effect of thermal fluctuations. After some validations steps in absence of fluctuations, we study the fluctuations in the particle's velocity at changing thermal energy,  applied force,  particle size, and particle wettability. The importance of fluctuations with respect to the mean settling velocity is quantitatively assessed, especially in comparison to unconfined situations. Results show that the expected effects of confinement are very well captured by the numerical simulations, wherein the confinement strongly enhances the importance of velocity fluctuations, which can be one order of magnitude larger than what expected in unconfined domains. The observed findings underscore the versatility of the proposed methodology in highlighting the effects of confinement on the motion of particles in presence of thermal fluctuations.
\end{abstract}

\maketitle

%%%%%%%%%%%%%%%%%%%%%%%%%%%%%%%%%%%%%%%%%%%%%%%%%%%%%%%%%%%%%%%
\section{Introduction}\label{sec:intro}
%%%%%%%%%%%%%%%%%%%%%%%%%%%%%%%%%%%%%%%%%%%%%%%%%%%%%%%%%%%%%%%
Complex flow phenomena involving dispersions of particles moving in viscous fluids are of interest for their theoretical relevance in the framework of non-equilibrium statistical mechanics~\cite{landau2013course,reichl1999modern}. Such phenomena are also relevant in a variety of applications, ranging from large~\cite{toschi2009lagrangian} to small scales~\cite{ladd2001lattice}. The corresponding theoretical description at the large scales hinges on the deterministic Navier-Stokes equations~\cite{lamb1993hydrodynamics,milne1996theoretical}, suitably coupled to the surface of the particles via hydrodynamic boundary conditions; these, in turn, account for the affinity of the particle towards the fluid and result in macroscopic properties, such as slip, wettability, etc. The deterministic dynamics of the Navier-Stokes equations is however \R{unsuitable} for the description at smaller scales, where the assumptions of negligible fluctuations cease to be valid. Consistently, fluctuations need to be taken into account, see~\cite{Landau, Zarate2006} for some reference textbooks and~\cite{croccolo2016non,giraudet2015slowing,giraudet2016confinement} (plus references therein) for some recent works on the topic. In these conditions, mesoscale methods represent methods of choice~\cite{succi2001lattice}. By definition, mesoscale modeling is constructed at scales which are intermediate between the large scales and the small scales; hence, a suitable coarse-graining allows to recover the hydrodynamical description based on the Navier-Stokes equations. Additionally, one can enrich the modeling with nanoscale features like thermal fluctuations~\cite{Landau,Zarate2006}. Among all mesoscale methods, we are interested in the lattice Boltzmann models (LBM). Over the last decades, LBM \R{have been} successfully used to model complex hydrodynamic phenomena at large scales, such as particle suspensions~\cite{ladd2001lattice,wu2010simulating,nguyen2002lubrication}, non-ideal fluids with phase transition and/or phase segregation~\cite{he1999lattice, liu2012three, reis2007lattice, chiappini2019, chiappini2018ligament, milan2018lattice}, polymer flows~\cite{ahlrichs1999simulation,ahlrichs1998lattice,berk2005lattice}, active matter~\cite{de2016lattice} just to cite some prominent examples. Especially in the last decade, there has been a boost to push the applicability of LBM simulations towards nano-scales via the inclusion of thermal fluctuations~\cite{Varnik11,Ladd,Adhikari2005,Dunweg07,Gross10, KaehlerWagner13}, designing the so-called {\it fluctuating lattice Boltzmann} methodology (FLBM). This methodology has been recently applied to the study of multicomponent fluids in the presence of thermal fluctuations~\cite{Belardinelli15,Belardinelli19} and also to study the effects of thermally excited capillary waves on the break-up properties of a thin liquid ligament~\cite{xue2018effects}. In this paper, we couple the FLBM with a wetted finite-size particle model~\cite{jansen2011bijels}. \R{This prompts the need of understanding how to choose the various tunable parameters in this complex system to obtain a stable numerical simulation; these parameters include the particle's resolution, particle's wettabilities, Shan-Chen forcing coupling coefficient, collision model relaxation time, fluids density ratio, gravity force, confinement ratio, and thermal energy. Some preliminary data on the particle diffusivity and mean square displacement without external driving forcing have been presented in~\cite{xue2020lattice}. Therefore, here we} focus on the quantitative assessment of the potentiality of such coupled methodology in modeling fluctuations of finite-size particles in the presence of confinement \R{and external driving forces}.  Specifically, we quantitatively characterize the motion of a spherical colloidal particle driven by a constant body force in a confined nanofluidic channel. The channel is filled with a fluctuating multicomponent mixture of two fluids. The results of numerical simulations are compared with the expectations of a simplified hydrodynamical-Langevin model for a finite-size particle comprising Gaussian noise and \R{effective friction} that accounts for the effects of confinement. We observe that numerical simulations capture very well the theoretical expectations at changing the various free parameters in the problem, i.e. the particle radius, the thermal energy, the driving force, and the particle wettability. In particular, the increasing importance of particle velocity fluctuations (with respect to its mean settling velocity) at increasing confinement is correctly modeled by the simulations. These numerical observations bear \R{non-straightforward methodological importance}, in view of the fact that simulations with FLBM cannot be granted a-priori the ``hydrodynamical limit''~\cite{Belardinelli15,Belardinelli19}, hence one has to verify a-posteriori if the outcome of simulations is well captured by hydrodynamical models.\\
The paper is organized as follows. In~\cref{sec:methodology}, we summarize the essential methodological aspects of the FLBM for multicomponent fluids and the coupling between particles and the multicomponent fluid. In~\cref{sec:Friction} we will present the set-up for the numerical simulations and we will present validation studies in the absence of thermal fluctuations, by comparing the settling velocity with previous experimental and numerical data. Results in the presence of thermal fluctuations will be presented in~\cref{sec:Fluctuations}. Conclusions will follow in~\cref{sec:conclusions}.
\section{Methodology}\label{sec:methodology}
To model the bulk fluid, we consider LBM that allow for the simulations of multicomponent mixture of two components in the presence of thermal fluctuations~\cite{Belardinelli15}. We additionally introduce finite-size particles via a suitable coupling between the particle and the multicomponent fluid~\cite{ladd1994numerical,aidun1998direct,jansen2011bijels}. The essential technical details of the LBM used here are briefly summarized, the interested reader can refer to the reference works~\cite{Belardinelli15,ladd1994numerical,aidun1998direct,jansen2011bijels} for more extensive technical coverage.\\
The multicomponent LBM considers the evolution equation of probability distribution functions, $f_{l i}(\mathbf{x},t)$, representing the probability density to find a particle of fluid component $l=A,B$ with kinetic velocity ${\bm c}_i$ in the space-time location $(\mathbf{x},t)$. Lattice velocities are discretized ($i=0,1,...,Q-1$) and we employ the D3Q19 model, with $Q=19$ velocity directions. The density of each component and the mixture velocity can be obtained via a proper coarse-graining in the kinetic velocity
\begin{equation}\label{eq:density}
\rho_{l}(\mathbf{x}, t) = \sum_{i=0}^{Q-1} f_{l i}(\mathbf{x}, t),  \hspace{.2in} \rho_{\st{tot}}(\mathbf{x}, t)\vv(\mathbf{x}, t) = \sum_{i=0}^{Q-1}\sum_{l=A, B} f_{l i}(\mathbf{x}, t)\mathbf{c}_{i}
\end{equation}
being $\rho_{\st{tot}}(\mathbf{x},t)=\sum_{l=A, B} \rho_{l}(\mathbf{x},t)$ the total density. The evolution equation for the distribution functions over a unitary time step is given by
\be\label{eq:lbe}
f_{l i}(\mathbf{x}+\mathbf{c}_{i},t+1)-f_{l i}(\mathbf{x}, t)=\mathfrak{L}\left[f_{l i}(\mathbf{x},t )-f^{(eq)}_{l i}(\mathbf{x},t )\right]+S^{(F)}_{l i}(\mathbf{x},t) + \xi_{l i}(\mathbf{x},t)  \hspace{.2in} l=A,B.\\
\ee
The collision operator $\mathfrak{L}$ is designed in such a way that it expresses the relaxation of the whole system towards a local Maxwellian distribution function $f^{(eq)}_{l i}(\mathbf{x},t)$~\cite{succi2001lattice, kruger2017lattice}. Technically, we make use of the MRT (multiple relaxation time) scheme~\cite{DHumieres02,Dunweg07,SchillerThesis}: the distribution functions are decomposed in modes (density, momentum, stress, etc) and the action of $\mathfrak{L}$ consists in relaxing the different modes with different relaxation times~\cite{DHumieres02}. The relaxation time of the momentum modes will determine the species diffusivity~\cite{Dunweg07}, whereas the relaxation time of the stress modes will determine the fluid viscosity~\cite{DHumieres02,Dunweg07}. The term $S^{(F)}_{l i}(\mathbf{x},t)$ is a deterministic source term, accounting for the external body forces and the interactions between the two components. For the modeling of non-ideal interactions, we adopt the Shan-Chen formulation for multicomponent mixtures~\cite{SC93,SC94,Zhang11,SbragagliaBelardinelli,SegaSbragaglia13}, where the force experienced by the fluid component $l$ due to the surrounding fluid component $l'$ can be written as 
\begin{equation}\label{sc-force_eq}
F_{l}(\mathbf{x},t) = - {\cal G} \rho_{l}(\mathbf{x},t) \sum_{l' \neq l}\sum_{i=0}^{Q-1} \omega_{i} \rho_{l'}(\mathbf{x}+\mathbf{c}_{i},t) \mathbf{c}_{i}  \hspace{.2in}
\end{equation}
where ${\cal G}$ is a strength coefficient and $\omega_{i}$ a suitable weight needed to impose the isotropy in the interactions~\cite{SC93,SC94,SbragagliaBelardinelli}. In all the simulations performed, we consider a non-ideal mixture with ${\cal G}= 1.5$ (lattice Boltzmann units, lbu) and we simulate a bulk fluid with a majority of component $A$ and fluid densities $\rho_A=2.21$ lbu (majority component) and $\rho_B=0.09$ lbu (minority component). The term $\xi_{l i}(\mathbf{x},t)$ is a stochastic force, which adds to the deterministic evolution a stochastic term. The stochastic terms are chosen in such a way that the conserved mass densities do not receive any stochastic force, while non-conserved modes receive a stochastic force in compliance with the fluctuation-dissipation relation~\cite{Belardinelli15}. The FLBM equations~\cref{eq:lbe} imply evolution equations for macroscopic density and velocity. If we apply a Chapman-Enskog procedure~\cite{Dunweg07,SchillerThesis} by treating the stochastic terms as ``generic'' forcing terms, the macroscopic equations of a binary mixture in the presence of thermal fluctuations are recovered for the fluid densities and the hydrodynamical velocity ${\bf v}^{(H)}={\bf v}+({\bf F}_{A}+{\bf F}_{B})/2 \rho_{\st{tot}}$ (superscript $T$ means transposition)~\cite{Zarate2006,Belardinelli15}  \footnote{When applying the Chapman-Enskog procedure to obtain the momentum equation, we can first consider the sum of all populations $\sum_{l} f_{l i}(\mathbf{x})$ as an ``effective'' 1 component population; then, we can rely on the known results for 1 component systems~\cite{Dunweg07,SchillerThesis}.}
\be
\partial_t \rho_{\st{tot}} + {\bf \nabla} \cdot (\rho_{\st{tot}} {\bf v}^{(H)})=0
\ee
\be
\partial_t (\rho_{\st{tot}} {\bf v}^{(H)})+{\bf \nabla} (\rho_{\st{tot}} {\bf v}^{(H)} {\bf v}^{(H)}) = -{\bf \nabla} P_b + {\bf \nabla} \cdot [\eta ({\bf \nabla} {\bf v}^{(H)} + ({\bf \nabla} {\bf v}^{(H)})^T )+{\bf \Sigma}_{\st{tens}}] + \rho_{\st{tot}} \fg
\ee
\be
\partial_t \rho_A+{\bf \nabla} \cdot (\rho_A {\bf v}^{(H)})={\bf \nabla} \cdot \left[ D {\bf \nabla} \mu + \bm{\Psi}_{\st{vec}} \right]
\ee
where the bulk pressure $P_b$ and the chemical potential $\mu$ assume the form  $P_b=\frac{1}{3} \rho_{\st{tot}} +\frac{{\cal G}}{3} \rho_A \rho_B$ and $\mu=\frac{1}{3} \log \rho_A - \frac{1}{3} \log \rho_B+\frac{1}{3} {\cal G} (\rho_A -\rho_B)$~\cite{Belardinelli15}, $\fg$ is external body force density acting on the fluid. The transport coefficients $D$ and $\eta$ are related to the relaxation times of the fluid. These will be fixed to $D=1/6$ lbu and $\eta=0.383$ lbu in all the simulations performed. The capital Greek symbols identify the stochastic stress (${\bf \Sigma}_{\st{tens}}$) and the stochastic diffusion ($\bm{\Psi}_{\st{vec}}$) contributions to the equations of hydrodynamics 
\be
{\bf \Sigma}_{\st{tens}}=\sqrt{\eta \kbt}({\bf W_{\st{tens}}}+{\bf W}_{\st{tens}}^T) \hspace{.2in} \bm{\Psi}_{\st{vec}}=\sqrt{2 D  \kbt} {\bf W_{\st{vec}}}
\ee
where $\kbt$ is the thermal energy, while ${\bf W}_{\st{tens}}$ and ${\bf W}_{\st{vec}}$ are a Gaussian tensor and a Gaussian vector with independent and uncorrelated components and variance equal to unity. In both homogeneous and heterogeneous systems, validations for the stochastic term have been done in previous studies~\cite{Belardinelli15,Belardinelli19}. Specifically, for homogeneous systems, it was demonstrated that the correlations of the hydrodynamical fields are exactly those that can be predicted from the fluctuating hydrodynamic equations that we write above. For heterogeneous systems, the model has been successfully benchmarked against capillary fluctuations at non-ideal interfaces. In a more recent study~\cite{Belardinelli19}, the model has also been shown to reproduce quantitative details of non-equilibrium fluctuations. We remark that the hydrodynamical equations reported above are obtained via a Chapman-Enskog procedure. This requires fields that slowly vary in time and space, hence in the presence of fluctuations it may become questionable. We will discuss more in details these issues while presenting the results of numerical simulations.\\
For the LBM modeling of the particle, we follow References~\cite{ladd1994numerical,aidun1998direct,jansen2011bijels}. The particle is modeled on the lattice, by declaring the fluid nodes belonging to the particle (``particle nodes''), as sketched in~\cref{fig:sketch_settling}. The motion of the particle is determined by Newton's equation~\cite{aidun1998direct}, and the evolution of the finite-size particle is solved with the leap-frog algorithm~\cite{allen2017computer}. The integration of the leap-frog algorithm has been validated in previous studies for the finite-size particles in turbulent channel flows~\cite{gupta2018simulation, gupta2018computational}. The bounce back boundary condition is implemented at the interface between the particle and the fluid~\cite{ladd1994numerical}. During the bounce back procedure, the particle exchanges the momentum with the surrounding fluid. Due to the particle movement in the fluid, there will be the creation of new particle nodes which originally were fluid nodes (cover-nodes behavior). Analogously, the movement of the particle can delete the particle nodes and create new fluid nodes (uncover-nodes behavior). In order to impose the total mass conservation, we implement the mass correction algorithm described in~\cite{jansen2011bijels}. Also, we introduce a virtual fluid layer~\cite{jansen2011bijels} at the interface between the particle and the fluid to be able to tune the particle's wettability. In such a layer, the fluid densities are set equal to the average densities of the neighboring fluid nodes, plus a correction $\Delta \rho$ that is instrumental to model the affinity of the particle towards the two components. The wettability properties described in the following (i.e. hydrophobic, neutral, hydrophilic) refer to the affinity of the particle towards the majority component in the bulk phase.

%%%%%%%%%%%%%%%%%%%%%%%%%%%%%%%%%%%%%%%%%%%%%%%%%%%%%%%%%%%%%%%%%%
\section{Numerical set-up and Validation}\label{sec:Friction}
\begin{figure}[H]
\centering
\includegraphics[width=0.75\textwidth]{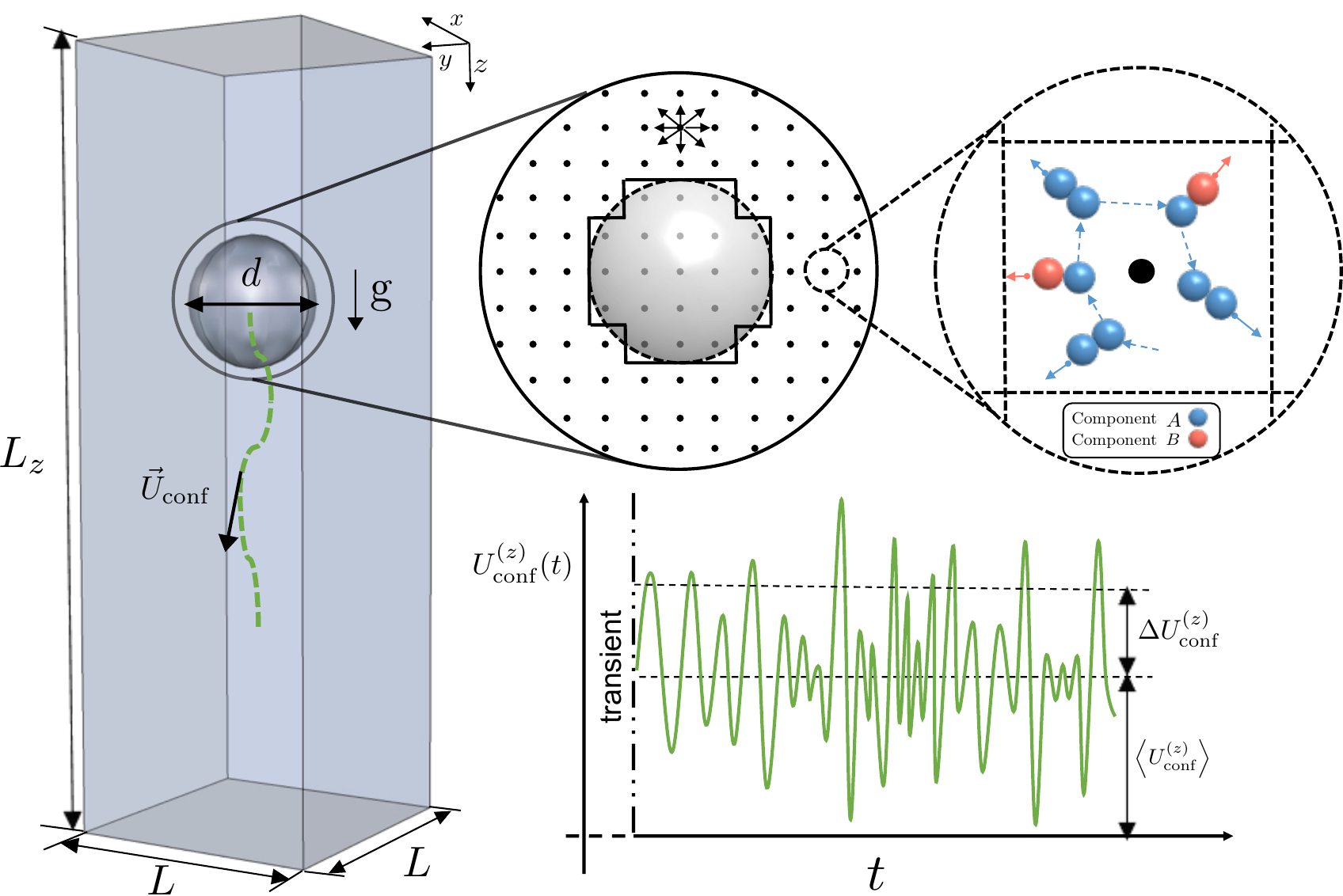}
\caption{Sketch of the setup for the particle settling numerical simulations. The computational box is a rectangular parallelepiped of height $L_z$ and square base $L \times L$. The solvent fluid is a fluctuating mixture of two non-ideal components, $A$ and $B$, with majority of the component $A$ in the bulk phase. The whole system is under the effect of the body force density, $\fg$, acting in the z direction. The mixture and the particle are simulated using lattice Boltzmann models (LBM) on a regular three dimensional lattice (cfr.~\cref{sec:methodology}). The LBM implementation is further equipped with thermal fluctuations (fluctuating LBM, FLBM~\cite{Belardinelli15}) to mimic the effect of noise at the small scales. The particle's velocity is tracked as a function of time and we quantify its statistical properties (average $\langle \uconf \rangle$ and fluctuations $\Delta \uconf$) in the statistically steady state.} \label{fig:sketch_settling}
\end{figure}

%%%%%%%%%%%%%%%%%%%%%%%%%%%%%%%%%%%%%%%%%%%%%%%%%%%%%%%%%%%%%%%%%%
The set-up for the numerical simulations is sketched in~\cref{fig:sketch_settling}. A particle with diameter $d$ is placed in a long channel with square cross section $L \times L$. The particle is initially placed with its center of mass lying in the center of the square cross-section and is driven by a body force density, $\fg$, acting in the z direction. The resulting force on the particle is
\be\label{eq:fp}
\fp=\frac{4}{3}\pi \left(\frac{d}{2}\right)^3(\rhop-\rho_{\st{tot}})\fg
\ee
where $\rhop$ is the particle density which is set to $\rhop=2\rho_{\st{tot}}$. The channel is resolved with $L \times L \times L_z$ = $60 \times 60 \times 900$ lbu. The channel is closed with walls in all directions; this choice is instrumental to fully appreciate the effects of confinement. A neutral wettability boundary condition is chosen for all the bounding walls, while three different wettabilities are considered at the interface between the particle and the fluid: these correspond to wetting angles $\theta=120.5$, $\theta=90^{\circ}$, $\theta=55.0$ and will be denoted hereafter as ``hydrophobic'', ``neutral'' and ``hydrophilic''. Due to the wide range of parameters, it is highly challenging to fully validate the theoretical prediction with numerical simulations. In this work, we spent around 1.4 Million computing hours to accomplish the validation. The square cross section is kept fixed in all numerical simulations, while different particle's diameters are considered. Different values of the thermal energy and driving force are also simulated. The simulations parameters are chosen in the following ranges: $d/L \in [0.133:0.67]$, $\kbt \in [1\cdot10^{-5}:0.45 \cdot 10^{-3}]$ lbu, $\fg \in [5 \cdot 10^{-7}: 5 \cdot 10^{-5}]$ lbu.\\
%%%%%%%%%%%%%%%%%%%%%% FIG 1 %%%%%%%%%%%%%%%%%%%%%%%%%%%%%%%%%%%%%
We have first validated the numerical set-up without thermal fluctuations. To this aim, we measured the steady settling velocity of the particle at changing the particle diameter. The steady settling velocity in the confined (conf) channel will be proportional to the driving force and inversely proportional to the friction $\gconf$ 
\be
\uconf=\frac{\fp}{\gconf}.
\ee
In unconfined (unconf) domains one would expect the Stokes-law for the friction $\go=3 \pi \eta d$. However, it is known from the literature that confinement enhances friction and reduces the settling velocity in comparison to the unconfined cases~\cite{happel2012low,ganatos1980strong,ganatos1980strong1,keh2001slow,miyamura1981experimental}. We therefore focused the attention on the ratio $\cm$ as a function of $d/L$. $\cm$ is defined as the ratio between the particle's settling velocity under confinement and the Stokes' prediction for an unconfined particle driven by the same body force
\be\label{eq:cm}
\cm=\frac{\uconf}{\uo}.
\ee
Notice that $\cm$ is a function of the aspect ratio $d/L$, with the property that $\lim_{d/L\rightarrow 0} \cm=1$. Results are in good agreement with the experimental observations. Discrepancies which may be observed with coarser grids (results from~\cite{aidun1998direct}) become essentially negligible with our finer grids. We also observe that there is almost no dependency on the wettability condition. 
%\MS{The following part is very difficult to understand: what do you want to say ? Maybe better to remove. I am not sure that wettability effects should not be observed...  {\it Since the particle is immersed in the multicomponent fluid, even if the particle is hydrophilic or hydrophobic to the majority or minority of the bulk, the neighbouring fluid nodes near the particle may more covered with one of the two components and less covered the other one. However, lattice nodes are overall conserved with total mass, so that the particle behaves similar to the single component case. That is the reason that we don't observe special confinement effect due to wettability}.}
%%%%%%%%% FIG 2%%%%%%%%%%%%%%
\begin{figure}[H]
\centering
\includegraphics[width=0.5\textwidth]{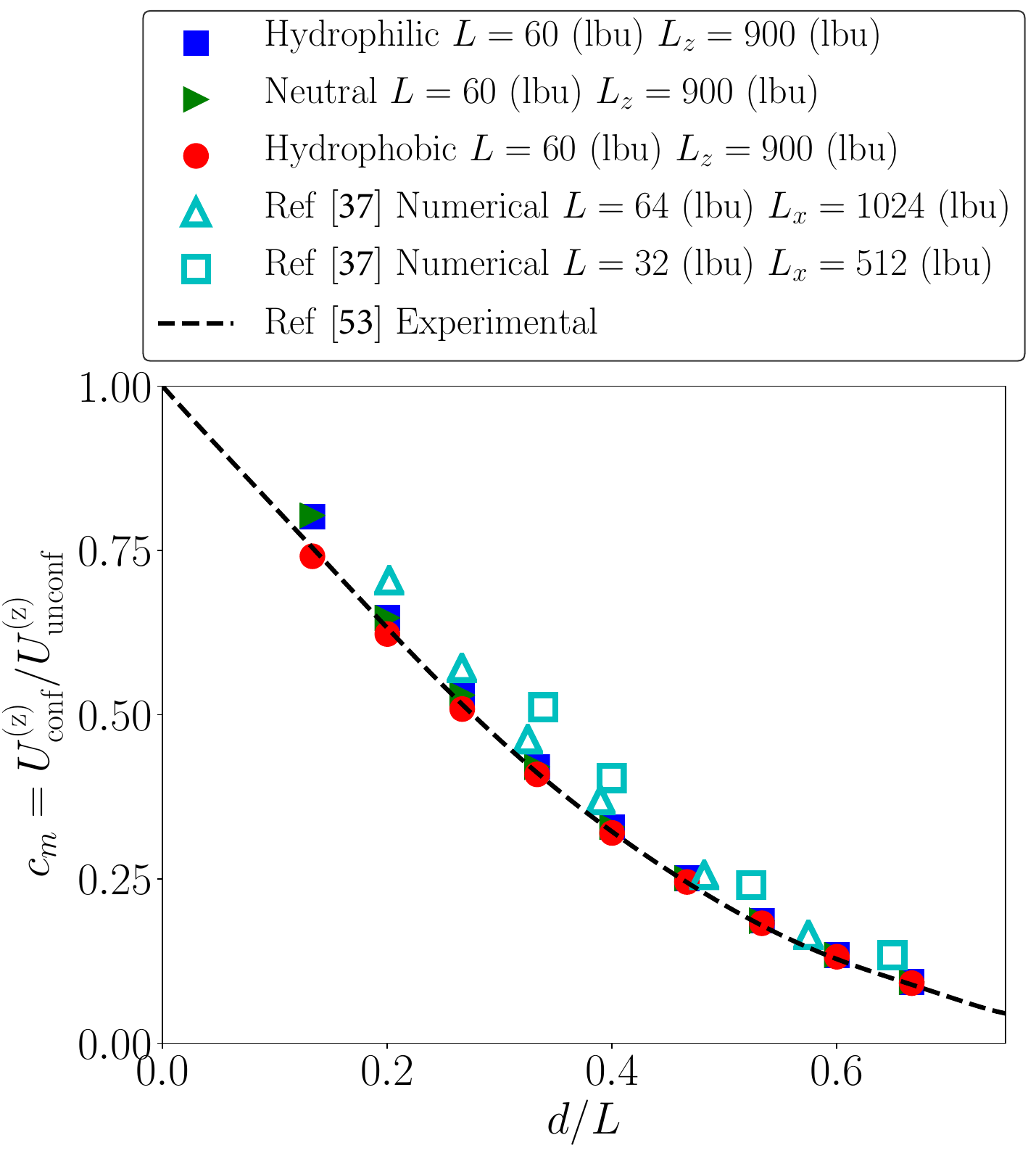}
\caption{We report the ratio, $\cm$, between the particle's settling velocity under confinement and the Stokes' prediction for an unconfined particle driven by the same body force (cfr.~\cref{eq:cm}). The ratio $\cm$ is considered as a function of the degree of confinement $d/L$ (cfr.~\cref{fig:sketch_settling}) and for different wettability boundary conditions at the particle's surface: hydrophilic (blue squares), neutral (green triangles), hydrophobic (red circles). Results are compared with the experimental results in~\cite{miyamura1981experimental} and with the numerical results in~\cite{aidun1998direct}. Our numerical investigation agrees well with the previous numerical and experimental results.}
\label{fig:particleSettlingAidun_setup}
\end{figure}
%%%%%%%%%%%%%%%%%%%%%%%%%%%%%%%%%%%%%%%%%%%%%%%%%%%%%%%%%%%%%%%%%%%
\section{Results and Discussions}\label{sec:Fluctuations}
\begin{figure}[H]
\centering
\includegraphics[width=1.0\textwidth]{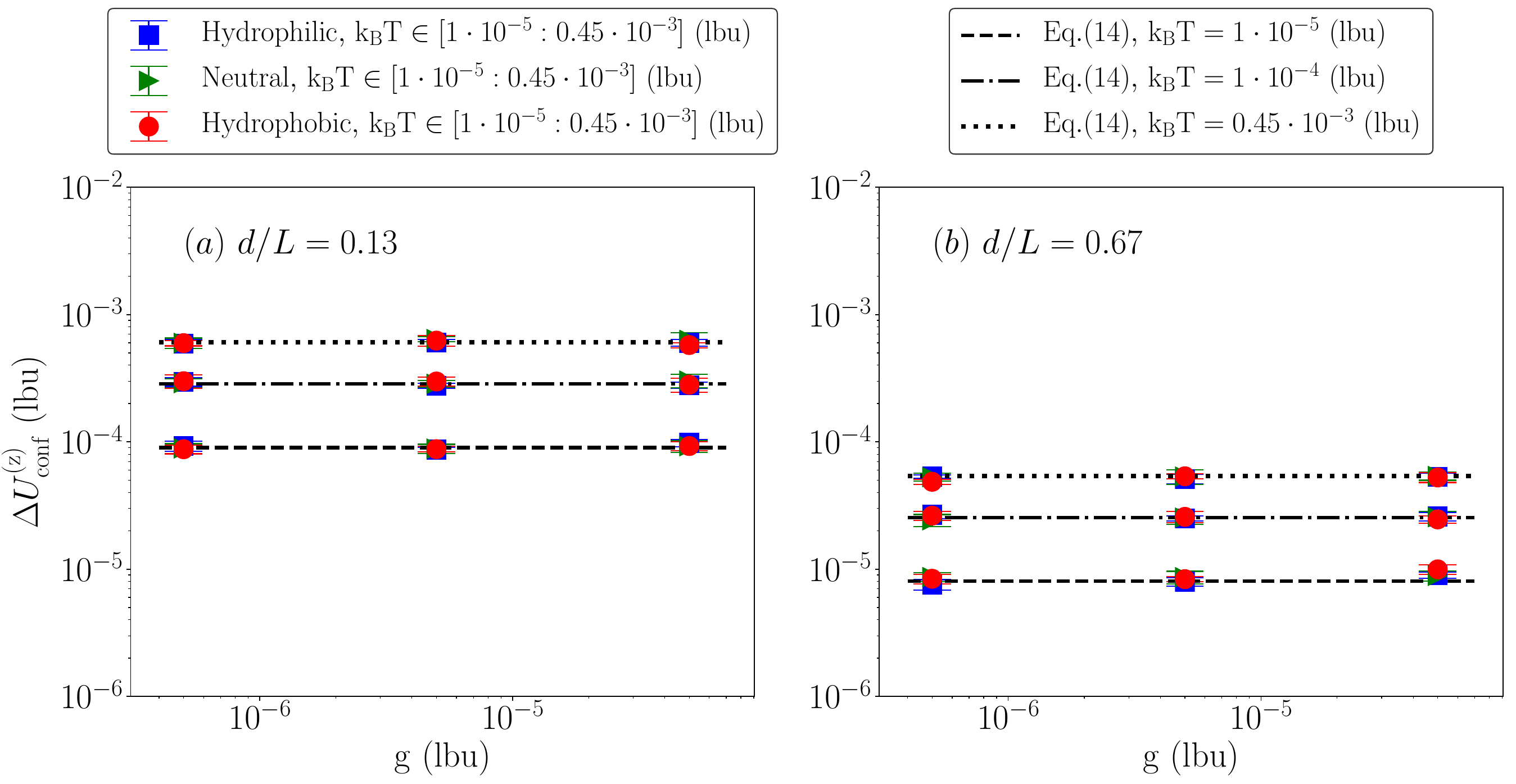}
\caption{Particle's velocity fluctuations $\Delta \uconf$ as a function of the body force density $\fg$, at changing $d/L$ and thermal energy $\kbt$. \R{Hydrophilic (blue squares), neutral (green triangles) and hydrophobic (red circles) cases have been shown.} Results are compared with the theoretical prediction given in~\cref{eq:FLUCTUATIONSSCALING}. Panel(a) shows the results under the degree of the confinement $d/L=0.13$, and Panel (b) presents the results at $d/L=0.47$. Our simulation data fit well with the theory at all $\fg$ for Panel(a) and Panel(b). Also, particle's velocity fluctuations show no dependency on the body force density $\fg$. When the particle reaches the stationary state, we equally split the data set in five time intervals. Error bars are the standard deviations from different groups of the configurations.}
\label{fig:duFtotal}
\end{figure}

\begin{figure}[H]
\centering
\includegraphics[width=1.0\textwidth]{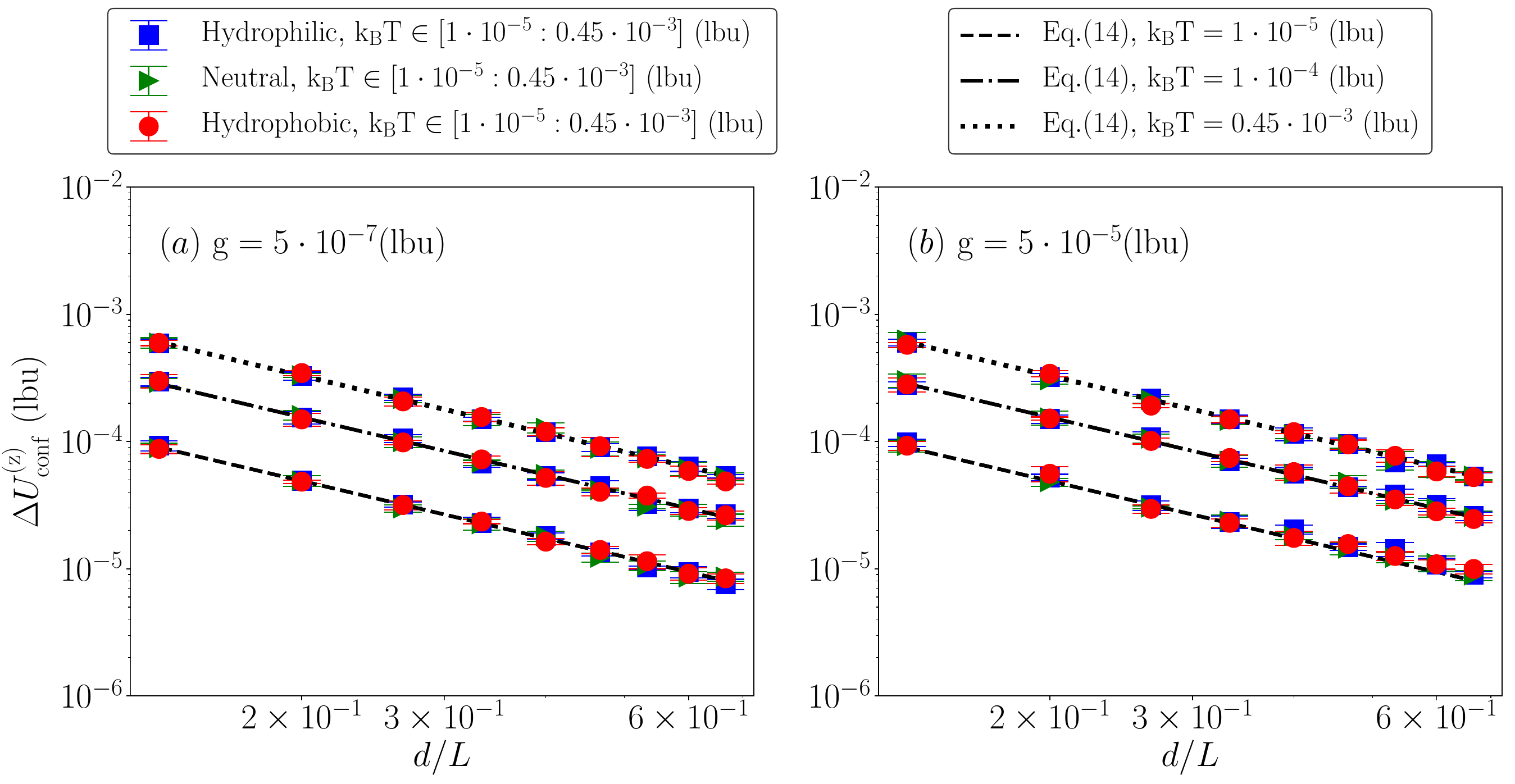}
\caption{Particle's velocity fluctuations as a function of $d/L$ at changing the body force density $\fg$ and the thermal energy $\kbt$. \R{Hydrophilic (blue squares), neutral (green triangles) and hydrophobic (red circles) cases have been shown.}  Results are compared with the theoretical prediction given in~\cref{eq:FLUCTUATIONSSCALING}. Panel(a) shows results at $\fg=5\cdot10^{-7}\mathrm{(lbu)}$, and Panel(b) shows results at the largest body force density $\fg=5\cdot10^{-5}\mathrm{(lbu)}$. Three different lines are the theoretical predictions from~\cref{eq:FLUCTUATIONSSCALING} at $\kbt=1\cdot10^{-5}, 1\cdot10^{-4}, 0.45\cdot10^{-3}\mathrm{(lbu)}$. Our simulation data fit well with the theory at all $d/L$. When the particle reaches the stationary state, we equally split the data set in five time intervals. Error bars are the standard deviations from different groups of the configurations. }
\label{fig:dudtotal}
\end{figure}

\begin{figure}[H]
\centering
\includegraphics[width=1.0\textwidth]{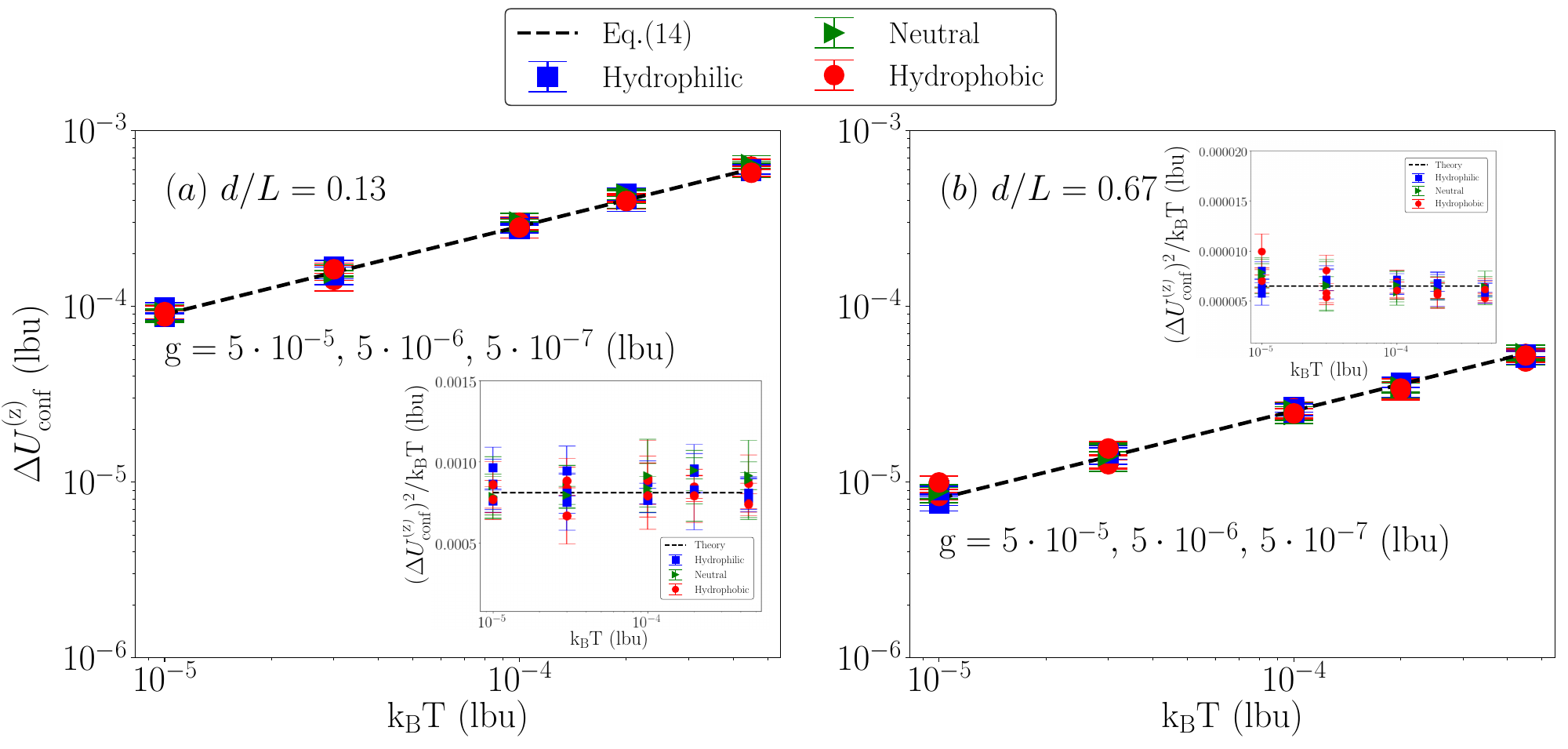}
\caption{Particle's velocity fluctuations as a function of the thermal energy $\kbt$ at changing the body force density $\fg$ and $d/L$. \R{Hydrophilic (blue squares), neutral (green triangles) and hydrophobic (red circles) cases have been shown.} The back dotted line is the theoretical predictions given in~\cref{eq:FLUCTUATIONSSCALING}. Panel (a) and (b) show that simulation data match well with the theory at all $\kbt$ under the body force density $\fg=5\cdot10^{-5}, 5\cdot10^{-6}, 5\cdot10^{-7}\mathrm{(lbu)}$. When the particle reaches the stationary state, we equally split the data set in five time intervals. Error bars are the standard deviations from different groups of the configurations. The subplots of Panel (a) and (b) checks the $(\Delta \uconf)^2/\kbt$, at changing $d/L$ the value remain constant which is equal to $1/\massp$.}
\label{fig:dukbttotal}
\end{figure}

\begin{figure}[H]
\centering
\includegraphics[width=1.0\textwidth]{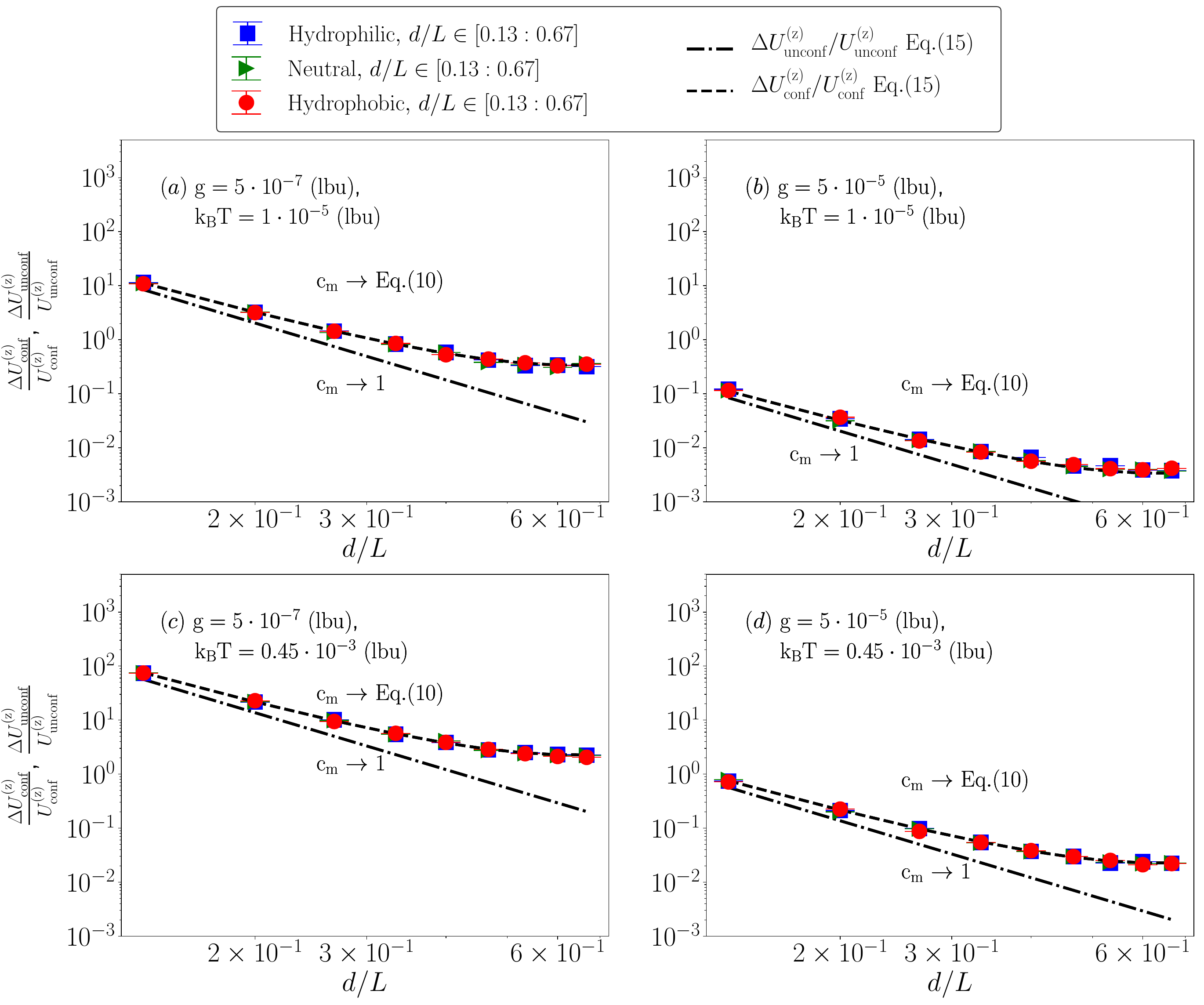}
\caption{Particle's velocity fluctuations normalized to the average velocity in both confined (conf) and unconfined (unconf) environments, as a function of the normalized particle's diameter $d/L$. \R{Hydrophilic (blue squares), neutral (green triangles) and hydrophobic (red circles) cases have been presented.} We change both the body force density $\fg$ and the thermal energy $\kbt$: $\fg=5 \cdot 10^{-7}$ lbu, $\kbt=1\cdot 10^{-5}$ lbu (Panel (a)), $\fg=5 \cdot 10^{-5}$ lbu, $\kbt=1\cdot 10^{-5}$ lbu (Panel (b)), $\fg=5 \cdot 10^{-7}$ lbu, $\kbt=0.45 \cdot 10^{-3}$ lbu (Panel (c)), $\fg=5 \cdot 10^{-5}$ lbu, $\kbt=0.45 \cdot 10^{-3}$ lbu (Panel (d)). Theoretical prediction for the confined cases is given in~\cref{eq:normalizedconf}. Theoretical prediction for unconfined cases is obtained using~\cref{eq:normalizedconf} with $\cm=1$. When the particle reaches the stationary state, we equally split the data set in five time intervals. Error bars are the standard deviations from different groups of the configurations.}
\label{fig:du_uc_F_kbt_d_total} 
\end{figure}

%%%%%%%%%%%%%%%%%%%%%%%%%%%%%%%%%%%%%%%%%%%%%%%%%%%%%%%%%%%%%%%%%
After the validation of the results in the absence of thermal fluctuations, we switched-on the thermal noise in the LBM simulations and studied the corresponding fluctuations in the particle's velocity $\uconf$ in the confined environment. As we have seen in the previous section, the friction acting on the particle is clearly affected by confinement, and it increases with respect to the unconfined case. This increase in friction is quantitatively well reproduced by the simulations (cfr.~\cref{fig:particleSettlingAidun_setup}). Fluctuations are added in the LBM in compliance with the fluctuation-dissipation balance~\cite{Belardinelli15,Belardinelli19}; thus - as a first guess - one could invoke a simplified picture based on a Langevin equation for the particle's velocity in the direction of the body force~\cite{risken1996fokker} 
\be\label{eq:langevin}
\massp \frac{d\uconf(t)}{dt} + \gconf \uconf(t) = F_{\mathrm{p}} + \zeta(t), 
\ee
where $\massp=\pi d^3 \rhop/6$ represents the particle's mass. The scalar term $\zeta(t)$ stands for the stochastic noise in compliance with the fluctuation dissipation theorem~\cite{risken1996fokker}, i.e.
$$
\langle \zeta(t) \zeta(t') \rangle = 2 \gconf \kbt \delta(t-t').
$$
Establishing the correspondence between the mesoscale FLBM dynamics (cfr.~\cref{sec:methodology}) and~\cref{eq:langevin} is not simple from the methodological point-of-view. Indeed, interpretations based on hydrodynamical equations for lattice Boltzmann simulations rely on a coarse-graining view in the kinetic velocity space and invoke some multi-scale expansion technique (e.g. Chapman-Enskog~\cite{succi2001lattice, Belardinelli15}) to find the corresponding hydrodynamical equations. By treating the stochastic source terms as ``generic'' one can surely carry out the detailed expansion calculations and derive fluctuating hydrodynamics equations (cfr.~\cref{sec:methodology}). It has to be noted, however, that such a procedure typically requires that fields under study slowly vary in time and space; thus, the ``equivalence'' between the FLBM simulations and the simple model~\cref{eq:langevin} could well fail. One is therefore left with the need of assessing a-posteriori the correctness of numerical simulations and if they match the predictions of~\cref{eq:langevin} without fitting parameters. Based on this view, we started to analyze the statistical properties in the particle's velocity.
%%%%%%%%%%%%%%%%%%%%%%%%%% FIG 3 %%%%%%%%%%%%%%%%%%%%%%%%%%%%%%%%%%%%%%
\begin{figure}[H]
\centering
\includegraphics[width=1.0\textwidth]{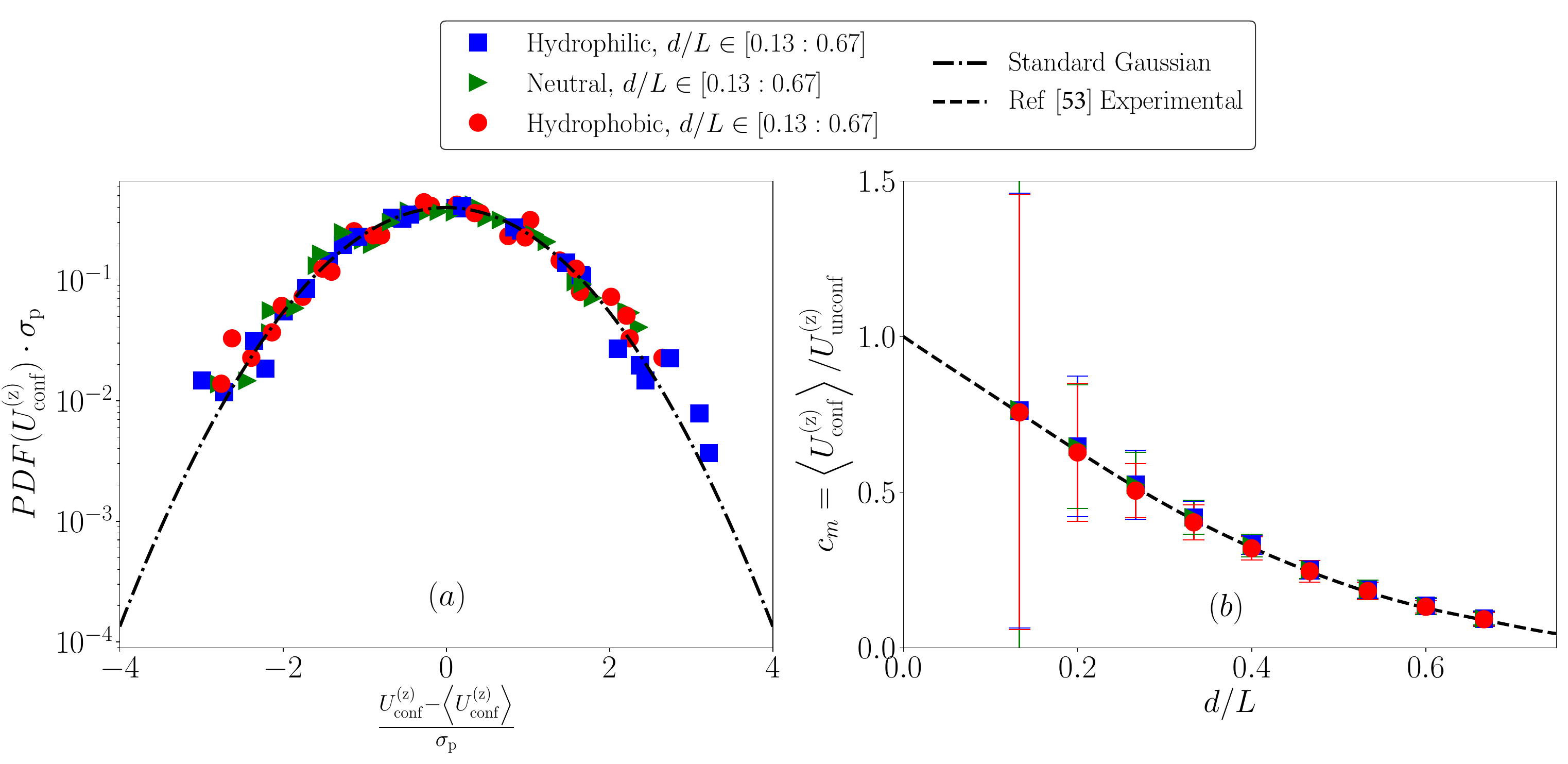}\label{fig:gaussianRescale}
\caption{Panel (a): Standardized PDFs of particle's settling velocity at fixed body force density $\fg=5\cdot 10^{-5}$ (lbu) and fixed thermal energy $\kbt=0.45\cdot 10^{-3}$ (lbu). Three different wettabilities were chosen: hydrophilic, neutral and hydrophobic. To make the PDFs comparable with a standard Gaussian distribution, we have considered rescaled variables with zero mean and unitary variance. Data come from different values of the particle diameter $d/L=0.13, 0.47, 0.67$. Results are matched well with standard Gaussian distribution. Panel (b): \R{we report $\cm$ as function of $d/L$ in three cases: hydrophilic, neutral and hydrophobic.} The quantity $\cm$ is computed as the ratio between the average particle's settling velocity under confinement and the Stokes' prediction for an unconfined particle driven by the same body force. Error bars are estimated from standard deviation of the particle's settling velocity fluctuations. }
\label{fig:velocity_PDF}
\end{figure}
%%%%%%%%%%%%%%%%%%%%%%%%%%%%%%%%%%%%%%%%%%%%%%%%%%%%%%%%%%%%%%%%%%%%%%
 The steady state predictions from~\cref{eq:langevin} imply a Gaussian distribution for the velocity fluctuations
\be\label{eq:PDF}
P(\uconf)=\sqrt{\frac{1}{2\pi \sigp^2}} e^{-\frac{(\uconf-\langle \uconf \rangle)^2}{2 \sigp^2}} \hspace{.2in} 
\ee
where
\be\label{eq:MEANVARIANCE}
\langle \uconf \rangle=\frac{F_{\mathrm{p}}}{\gconf}  \hspace{.2in}  \sigp^2=\frac{\kbt}{\massp}.
\ee
First of all we checked that $\langle \uconf \rangle=\frac{F_{\mathrm{p}}}{\gconf}$ holds and that the results are compatible with a Gaussian shape. We report in~\cref{fig:velocity_PDF} some representative results for different $d/L$ and different wettabilities, while keeping the body force density and the thermal energy fixed to $\fg=5\cdot 10^{-5}$ lbu and $\kbt=0.45\cdot 10^{-3}$ lbu. To check for the Gaussian shape, we report the PDF of the quantity $x=(\uconf-\langle \uconf \rangle)/\sigp$. As can be seen,  $\langle \uconf \rangle=\frac{F_{\mathrm{p}}}{\gconf}$ holds and the numerical results collapse well on the Gaussian shape $f(x)=e^{-x^2/2}/\sqrt{2 \pi}$. Then, we proceeded in characterizing the dependency of the particle's velocity fluctuations $\Delta \uconf=\sqrt{\sigp^2}$ on the three parameters $d/L$, $\kbt$ and $\fp$.  From~\cref{eq:MEANVARIANCE} and  $\massp=\pi d^3 \rhop/6$ one gets
\be\label{eq:FLUCTUATIONSSCALING}
\Delta \uconf= \sqrt{\frac{6}{\pi \rhop L^3}}\,(\kbt)^{1/2}\,(d/L)^{-3/2}.
\ee
The behavior of the velocity fluctuations at changing $\fg$, $d/L$ and $\kbt$, is analyzed in~\cref{fig:duFtotal,fig:dudtotal,fig:dukbttotal}. 
%%%%%%%%%%%%%%%%%%%%%%%% FIG 4 %%%%%%%%%%%%%%%%%%%%%%%%%%%%%%%%%%%%%%%%%
The results are also compared with the prediction of~\cref{eq:FLUCTUATIONSSCALING}. As predicted by~\cref{eq:FLUCTUATIONSSCALING}, the velocity fluctuations are independent of the body force for fixed $d/L$ and $\kbt$ (cfr.~\cref{fig:duFtotal}); we also observe the scaling $\sim (d/L)^{-3/2}$ for fixed $\fg$ and $\kbt$ (cfr.~\cref{fig:dudtotal}) and the scaling $\sim (\kbt)^{1/2}$ for fixed $d/L$ and $\fg$ (cfr.~\cref{fig:dukbttotal}). To be noticed that not only the scaling laws but also the pre-factor $\sqrt{6/(\pi \rhop L^3)}$ in~\cref{eq:FLUCTUATIONSSCALING} matches very well with the numerical observations. \R{Overall, hydrophilic (blue squares), neutral (green triangles), and hydrophobic (red circles) results are well overlapping. we observe little dependency on the particle's wettability. }
%Moreover, when the force is $\fg=5\cdot 10^{-5}$ lbu and the thermal energy is $\kbt=0.45\cdot 10^{-5}$, the numerically evaluated $\Delta \uconf$ shows deviations from the theory. These deviation points are due to the fact that some numerical errors are introduced by the cover-uncover behavior (cfr. \cref{sec:methodology}), and these numerical errors are large for large forcing cases. Thus, in the simulations performed with the largest forcing and smallest $\kbt$, the numerical errors can overcome the (physical) stochastic contribution induced by $\kbt$ and produce deviations with respect to the theoretical predictions.
%%%%%%%%%%%%%%%%%%%%%%%% FIG 5 %%%%%%%%%%%%%%%%%%%%%%%%%%%%%%%%%%%%%%%%%
%%%%%%%%%%%%%%%%%%%%%%%% FIG 6 %%%%%%%%%%%%%%%%%%%%%%%%%%%%%%%%%%%%%%%%%
%%%%%%%%%%%%%%%%%%%%%%%%%%%%%%%%%%%%%%%%%%%%%%%%%%%%%%%
Finally, in~\cref{fig:du_uc_F_kbt_d_total} we consider velocity fluctuations normalized to the mean settling velocity \R{in hydrophilic, neutral and hydrophobic cases}, to highlight the impact of the fluctuations with respect to the characteristic order magnitude of the velocity. Based on~\cref{eq:FLUCTUATIONSSCALING,eq:cm} and $\uo=\fp/\go$, $\go=3 \pi \eta d$, we obtain
\be\label{eq:normalizedconf}
\frac{\Delta \uconf}{\uconf} = 18 \sqrt{\frac{6 \eta^2}{(\rho_A+\rho_B)^2 \pi \rhop L^{7}}}\frac{ \,(\kbt)^{1/2} \,(d/L)^{-7/2}\,\fg^{-1}}{\cm}
\ee
where we have related the particle's velocity to the unconfined velocity via the ratio $\cm$ (cfr.~\cref{eq:cm}). To gain insight on the importance of confinement, we also compared the present results with the unconfined predictions $\Delta \uo/\uo$ obtained by setting $\cm=1$ in \cref{eq:normalizedconf}. 
%%%%%%%%%%%%%%%%%%%%%%%%%%% FIG 7 %%%%%%%%%%%%%%%%%%%%%%%%%%%%%%%%%%%%%%%%%%%
In~\cref{fig:du_uc_F_kbt_d_total}, we report $\Delta \uconf/\uconf$, $\Delta \uo/\uo$ at changing $d/L$ for selected values of $\fg$ and $\kbt$ \R{in hydrophilic, neutral and hydrophobic cases}. We also compare with the theoretical predictions obtained from~\cref{eq:normalizedconf}. The numerical data are well in agreement with the theory for all values of $d/L$.  The unconfined theory is well reproduced only at small $d/L$, as expected. Notice that for the largest $d/L$ we observe a dramatic enhance of the importance of confinement, which is about one magnitude higher than the unconfined theory. This is expected based on the solution of~\cref{eq:MEANVARIANCE}.\\ 
Before closing this section, we stress once more that all the theoretical predictions that we have verified in the numerical simulations are the natural consequence of the simplified Langevin equation~\cref{eq:langevin}, where the noise follows the fluctuation dissipation theorem~\cite{risken1996fokker} and the friction accounts for confinement. The theoretical outcomes of such scenario predict that the mean settling velocity is reduced by confinement (cfr.~\cref{fig:particleSettlingAidun_setup}) and the fluctuations around the mean settling velocity are unchanged and equal to $\frac{\kbt}{\massp}$ (cfr.~\cref{eq:MEANVARIANCE}). What we target here is not the solution of such equation, but the verification that the results of the adopted methodology can be explained in the hydrodynamical limit with such an equation.
%%%%%%%%%%%%%%%%%%%%%%%%%%%%%%%%%%%%%%%%%%%%%%%%%%%%%%
\section{Conclusions}\label{sec:conclusions}
%%%%%%%%%%%%%%%%%%%%%%%%%%%%%%%%%%%%%%%%%%%%%%%%%%%%%
We studied the settling of a spherical particle with diameter $d$ in a fluctuating multicomponent fluid. The system is driven by a constant body force in a confined channel with a square cross-sectional area $L \times L$. Our simulations hinge on the fluctuating lattice Boltzmann methodology (FLBM) coupled with a finite-size particle model with tunable wettability~\cite{jansen2011bijels}. This methodological coupling has never been tested in the literature: due to the fluctuating nature of the lattice Boltzmann populations, it requires careful numerical verification in the assessment of its hydrodynamical properties.  We have first validated the numerical set-up in the absence of thermal fluctuations. In agreement with earlier numerical studies~\cite{aidun1998direct} on single component LBM, our numerical simulations with the multicomponent LBM well reproduce the frictional properties of a confined particle~\cite{miyamura1981experimental}. We have then switched-on thermal fluctuations, and we have systematically characterized the steady-state statistical properties (i.e. average and fluctuations) of the particle's velocity at changing the thermal energy $\kbt$, the degree of confinement $d/L$, the body force. The results of the numerical simulations show a neat matching with the predictions of a simplified Langevin-type scenario, accounting for the motion of a particle subject to the linear frictional law induced by confinement and in the presence of a stochastic force satisfying the fluctuation-dissipation theorem~\cite{kubo1966fluctuation}. We think this is a non-straightforward result since the coarse-grained description of FLBM requires some "hydrodynamical assumption", and this could be well violated by the presence of mesoscale fields that do not vary smoothly in space and time. On a quantitative basis, results in the presence of confinement show that the numerical tool is quite versatile in handling quantitative changes in frictional properties across orders of magnitude. Correspondingly, the measured ratio between the velocity fluctuations and the mean velocity comes out to be dramatically increased in the presence of confinement. Taken all together, the ``ensemble'' of simulations here proposed underscore the robustness and versatility of the proposed methodology in concrete applications involving the motion of colloidal particles in presence of confinement and multicomponent fluids.\\
\R{In future work, it would be interesting to explore regimes where noise effects produce a Brownian time larger than the Stokes' time. This would allow also to study lubrication effects coming from particle/wall interactions.}
\R{Another follow-up could be represented by the numerical simulations of the particle motion settling at the interface separating two immiscible fluids~\cite{boniello2015brownian}, where a change of wettability is expected to lead to more sizeable effects than those observed in the present study}. This makes the presented numerical results particularly relevant on the future perspective of achieving a further upgrade of the FLBM simulations as quantitative tools for the study of complex fluids with colloidal particles.

\section{ACKNOWLEDGEMENTS}
The authors would like to kindly acknowledge funding from the European Union's Horizon 2020 research and innovation programme under the Marie Sk\l{}odowska-Curie grant agreement No 642069 (European Joint Doctorate Programme ``HPC-LEAP"). This work is also supported by the European Research Council (ERC) under the European Union's Horizon 2020 research and innovation programme (Grant Agreement No. 882340). X.Xue acknowledges fruitful discussions and exchanges with A. Gupta in the early stage of the work.
\section{Author contribution statement}
All of the authors were involved in the preparation of the manuscript and have read and approved the final manuscript version.

\bibliographystyle{spphys}
\bibliography{prex}
\end{document}